\documentclass[aps,amsmath,amssymb,showpacs,twocolumn]{revtex4}
\usepackage{amsfonts}
\newcommand{\om}{\omega}

\newcommand{\Om}{\Omega}
\newcommand{\bk}{\bf k}
\newcommand{\bq}{\bf q}

\newcommand{\la}{\lambda}
\newcommand{\elTOph}{el{\textendash}TO{\textendash}ph}

\begin{document}

\title{ Contribution of interband effects caused by long-wavelength transverse \\
optical phonons to electron-phonon coupling in doped polar insulators}

\author{Aleksandr Pishtshev}
\email[]{E-mail: ap@eeter.fi.tartu.ee}
\affiliation{Institute of Physics, University of Tartu, Riia 142, 51014 Tartu, Estonia}

\begin{abstract}
\noindent
We estimate the contribution of the long-wavelength {\elTOph} interaction and discuss
the effect it has on electron pairing in doped polar systems like ${\rm SrTiO_3}$
and $\rm{PbTe}$. The analytical and numerical results presented in the study indicate that
the {\elTOph} interaction tends to contribute little to the total strength
of electron-phonon coupling in these and related materials. To explain this fact we consider
possible reasons why the effect of the polar long-wavelength transverse optical phonons
on the coupling constant $\la$ is far less than one might suppose.
\end{abstract}

\pacs{ 63.20.kd; 74.20.Fg; 74.70.-b; 77.84.Bw }

\maketitle

%
\section{Introduction}
A classical representative of polar crystals ${\rm SrTiO_3}$, which is the prototypical
among insulating $\rm ABO_{3}$ perovskite oxides, is a well-known material
with exceptional dielectric properties that made its attractive for a number
of electronic applications.
The ion-covalent bonds, the long-range dipole forces and extremely close proximity
to the ferroelectric state determine $\rm SrTiO_3$ as a model system with a great
potential for scientific investigations.
The electron-doped $\rm SrTiO_3$ represents an example
of a low carrier density system with a metallic-like conductivity.
Doping into insulating state of $\rm SrTiO_3$ is performed by electron donating ions
like ${\rm La}^{3+}$ for ${\rm Sr}^{2+}$ or ${\rm Nb}^{5+}$ for ${\rm Ti}^{4+}$
which transfer the doped electrons into the originally empty ${\rm d}^{0}$
configuration~\cite{Ohta1}.
Experimentally observable superconductivity
in the electron-doped $\rm SrTiO_3$ has been assumed to be described
by the plasmon-polar optic phonon mechanism~\cite{Takada1}.

The polar-coupling electron{\textendash}phonon system in $\rm SrTiO_3$ has attracted
a substantial amount of interest~\cite{Takada1,Kristoffel1,Mechelen,Devreese,Meevasana}.
The most important reason for this is that in such materials there exists a strong
coupling between electronic and lattice degrees of freedom. Correspondingly, 
the properties of conductivity electrons are strongly affected by local lattice properties.
The main motivation for the present study is related to previous work, in which we showed
that in polar crystals like $\rm SrTiO_3$ the strength of the interactions between valence
electrons and transverse optical (TO) phonon modes is especially
large~\cite{Pishtshev2,Pishtshev1}.
Application of these results to the case of doping will allow us to have a better
understanding of how electron dynamics is affected by the strong {\elTOph} interaction.
Note also that a good knowledge
about the roles of the coexisting interactions responsible for various physical
properties is valuable to further advances in polar and related systems. 
Therefore, it would be desirable to examine in detail whether and under which conditions
does the {\elTOph} interaction contribute to the total electron-phonon coupling constant.
The accomplishment of this task is the scope of the present communication.
%
\section{Model}
Consider a linearly coupled system of electrons and polar (infrared-active) TO lattice vibrations.
Assume that the band structure is represented by the set of one-particle Bloch wave functions
$\vert\,{\sigma}{\bf k}{\!\!}>$ and energies $E_{\sigma}({\bf k})$ which can be determined within
the framework of DFT-LDA methods with the correct accounting for the quasiparticle effects. 
The relevant Hamiltonian, which describes the dynamic mixing of electronic
states of the valence (${\sigma}=1)$ and conduction band (${\sigma}=2)$
caused by the TO phonons, reads
\begin{equation}\label{hamiltonian_el-ph}
H_{el-ph} =
{N}^{-1/2} \,
{\sum_{\sigma,\, \sigma^{'},\,j}}\,{\sum_{\bf k,\bf q}}\,
g_{{\sigma}{\sigma}^{'}}({\bf q}j) \,
a^+_{\sigma{\bf k}}a_{\sigma^{'}{\bf k-q}} \, u_{{\bf q}j} \,
\end{equation}
where the quantities $g_{{\sigma}{\sigma}^{'}}({\bf q}j)$ describe the {\elTOph} interaction,
$a^+$ ($a$) are the creation (annihilation) operators for the states
$\vert\,{\sigma}{\bf k}{\!\!}>{\rvert}_{{\sigma}=1,2}\,$,
$N$ is the number of unit cells;
the TO phonon modes are characterized by the normal coordinate $u_{{\bf q}j}\,$,
the wave vector ${\bf q}$ and the vibration branch $j\,$.

The consistent microscopic analysis of the model~(\ref{hamiltonian_el-ph}) is presented
in~\cite{Pishtshev2}. Some relevant characteristics of this model are summarized as follows:
(1) Eq.~(\ref{hamiltonian_el-ph}) is mostly related to polar crystals in which the difference
in long-range fields given by longitudinal and transverse optical phonon modes is
essential~\cite{Zhong}.
(2) For $q=0$, the interaction of band electrons and TO lattice vibrations
has completely the interband nature (the intraband (${{\sigma}={\sigma}^{'}}$) terms
in Eq.~(\ref{hamiltonian_el-ph}) vanish due to inversion symmetry~\cite{Kristoffel1,Pishtshev2}).
(3) The model represents the Fr\"{o}hlich-type
interaction which may give rise to an asymmetric charge distribution~\cite{Pishtshev2}.
(4) At the macroscopic level, the strength of the {\elTOph} interaction
is the direct effect of coupling between the dipole polarization associated with
the polar TO mode~\cite{BornM} and the resulting shift of the electron density~\cite{Pishtshev2}.
Note also that the electronic and structural significance of the {\elTOph} interaction
is incorporated within the vibronic model (e.g.,~\cite{Kristoffel1}).
An example here can be a family of the ferroelectric $\rm ABO_{3}$ perovskite oxides, in which 
hybridization~(\ref{hamiltonian_el-ph})
involves strong dynamic mixing between the $\rm O$ $\rm 2p$ and the ${\rm d}^{0}$
(${\rm Ti}^{4+}$, ${\rm Nb}^{5+}$, ${\rm Ta}^{5+}$, etc.) states caused by
the TO ${\rm F}_{1u}$ soft vibrations~\cite{Kristoffel2,Konsin3}.
Estimates of the {\elTOph} interaction constants
for a wide number of polar compounds are reported in~\cite{Pishtshev1}.
%
\section{Theoretical analysis in the long-wavelength limit}
A widely-used approach for estimating the electron-phonon coupling strength
is formulated in terms of the quasiparticle mass enhancement, $m^{*}/m$, due to the
electron-phonon renormalization factor $1+{\la}$ (e.g.,~\cite{Rickayzen,highTcBook}):
\begin{equation}\label{mass1}
m^{*}/m=1+{\la} \, .
\end{equation}
Here $\la$ is a dimensionless effective constant which characterizes the total strength
of electron-phonon coupling.
As well-known~\cite{Rickayzen,highTcBook}, transition temperatures in
conventional superconductors depend directly on $\la\,$.
According to common definitions (the representation of $\la$ as a sum over mode coupling constants)
one can define the partial coupling constant ${\la}_{\rm TO}$
which is related to the corresponding contribution of the {\elTOph} interaction to $\la\,$.

Being based on the model~(\ref{hamiltonian_el-ph}), our goal is to determine 
which characteristic values of
${\la}_{\rm TO}(j)$ for the $j$th TO mode correspond to the {\elTOph} interactions
(which are most strong in a polar crystal).
The simplest and practical method of obtaining ${\la}_{\rm TO}(j)$ is to calculate 
the corresponding corrections to the single-particle energies.
In the long-wavelength limit, upon comparing different physical processes and energy scales
operative in the system, we can conceptually isolate, in specific properties,
those effects arising from interactions between electrons and phonons from those
caused by the other factors.
This suggests ${\la}_{\rm TO}$ to be composed of three distinct elements, i.e. under the above
assumptions one has in the low-doping regime:
\begin{equation}\label{lambda2}
{\la}_{\rm TO}(j) = {(4N)}^{-1} \, {\sum_{\bq}}\,
{\cal V}(j,q) \times {\cal A}(j,q) \times {\cal K}(q) 
\end{equation}
where
\begin{align}\label{lambda3}
{\cal V}(j,q)= {\phantom{+}} & \frac{4{\vert \, g({\bq}j) \vert}^2}
{ {M_{j}{\om}_{j\bq}^{2}} \,\left(\, \Delta({\bk},{\bq})-{\hbar}{\om}_{j\bq} \right) }
\, {\Big \rvert}_{\bk =0} \, , \\ 
{\cal A}(j,q)= {\phantom{+}} &
\frac{ {\hbar}{\om}_{j\bq} }{ \Delta({\bk},{\bq})-{\hbar}{\om}_{j\bq} }\, 
{\Big \rvert}_{\bk =0}  \, , \\
{\cal K}(q) = \frac{m_{2}}{{\hbar}^{2}} & \Bigg\lbrack
\frac{\partial^{2}\Delta({\bk},{\bq})}{\partial \vert \bk \vert^{2}} -
\frac{2 \left( {\partial\Delta ( {\bk},{\bq})}/{\partial \bk } \right)^{2} }
{\Delta({\bk},{\bq})} \Bigg\rbrack
{\Big \rvert}_{\bk =0} \, , 
\end{align}
$\Delta({\bk},{\bq})={E_{2}({\bk})-E_{1}({\bk}+{\bq})}\,$,
$g({\bq}j) \equiv g_{{1}{2}}({\bf q}j)\,$, and
$m_{2}$ denotes the band mass in the absence of the {\elTOph} interaction.
The square of the TO phonon frequency $\Omega_{{\bf q}j}^{2}$
is represented as a sum of the unperturbed term ${\omega}^{2}_{{\bf q}j}$ and
the additional term $\Delta\omega^{2}_{{\bf q}j}$ which accounts for the {\elTOph} interaction:
$
\Omega^{2}_{{\bf q}j} = {\omega}^{2}_{{\bf q}j} + \Delta\omega^{2}_{{\bf q}j} \,.
$
We expect the {\textquotedblleft}normal{\textquotedblright} part ${\omega}_{{\bf q}j}$,
which may be regarded as the bare TO phonon frequency
with respect to the {\elTOph} interaction, to be free of softening anomalies.
The proposed analytical decomposition, Eq.~(\ref{lambda2}), allows us, firstly,
to distinguish various physical factors contributed to
${\la}_{\rm TO}(j)\,$, secondly, to treat them separately in order to understand
how the interplay of these elements affects the magnitude of ${\la}_{\rm TO}(j)\,$, 
and, thirdly, by using available experimental data, to obtain
reliable upper bounds for ${\la}_{\rm TO}(j)\,$.

The first element ${\cal V}(j,q)$ is the main player of the decomposition since
involves the effects of the hybridization of the electronic states by the TO phonon modes.
It can be analyzed by using our previous results~\cite{Pishtshev1,Pishtshev2}
as follows.
First of all, note that the quantity ${\cal V}(j,q)$ supplies a measure of the zone-center
TO phonon softening arising from the interband scatterings of the band electrons.
That is, ${\cal V}(j,q)$ is proportional to a ratio
of the electronic contribution ${\vert \, g({\bq}j) \vert}^2/{\Delta({\bq})}$,
determined by the electron-ion potential and the band structure,
and the lattice force constant ${M_{j}{\om}_{j\bq}^{2}}$ associated with
the bare TO vibrational mode.
The momentum dependence of the quantity $g({\bq}j)$ is seen from
the first-principles analysis~\cite{Pishtshev2}:
first, in a polar crystal, the maximum of $g({\bq}j)$ as a function of ${\bq}$
is strictly related to the center point ${\bq}=0\,$; second, 
the function $g({\bq}j)$ is decreasing with increasing ${\bq}\,$.
The last key aspect, which helps us to evaluate ${\cal V}(j,q)\,$, is 
the delicate balance of long- and short-range forces peculiar to ferroelectrics
(e.g.,~\cite{Blinc}). By using the connection between long-range forces and
the {\elTOph} interaction~\cite{Pishtshev1,Pishtshev2}, one can relate
the quantity $g({\bq}j)$ at ${\bf q}\rightarrow0$ to the relevant
material parameters.
As a result, in the long-wavelength limit, an upper-bound estimate for ${\cal V}(j,q)$
can be described by the expression:
\begin{equation}\label{lambda3A}
{\cal V}(j,q) \leq {\tau}_j \, , \quad
{\tau}_j = \frac { \widetilde{S}(j) } { \widetilde{S}(j) + {\epsilon}_{\infty}+2 } \, .
\end{equation}
Here,
$\widetilde{S}(j)=S(j)/{\Om}_{j0}^{2}$ and $S(j)$ is a dipole oscillator strength
associated with the $j$th zone-centre TO vibrational mode;
${\epsilon}_{\infty}$ is the core (electronic high-frequency) contribution
to the dielectric function. In the derivation of Eq.~(\ref{lambda3A})
we supposed that contributions of the zone-centre TO phonons to the dielectric function
${\epsilon}({\omega})$ of the far-IR spectral range are represented by a classical 
oscillator model~\cite{Kamba,Cowley2}:
\begin{equation}\label{dm-eq1} 
{\epsilon}({\omega}) \,=\, {\epsilon}_{\infty} \,+\,
{\sum_{j}}\, 
\frac {S(j)} {\, \Omega^{2}_{j0}{\,-\,}{\omega^{2}} \, } \,.
\end{equation}
It is seen that the parameter ${\tau}_j$ always less than unity so that
the inequality ${\cal V}(j,q)<1$ can be viewed as a reliable qualitative estimate.
Note that, in a macroscopic sense,
the difference between ${\rm 1}$ and ${\tau}_j$ can be considered as a measure
of the polarity in such a manner that in polar materials, such as ferroelectrics,
$1-{\tau}_j$ is much closer to zero than in compounds of lower polar nature. 

Following~\cite{Pishtshev1,Pishtshev2}, we note that the second element ${\cal A}(j,q)$
serves as an adiabatic parameter which relates two different energy scales {\textendash}
one is the electronic scale determined by magnitude of an insulating gap $E_{g}$;
the other is the conventional phonon frequency scale ${\hbar}{\om}_{j}$.
In view of the balance of long- and short-range forces as well as weak dispersion 
of the long-wavelength TO lattice vibrational modes,
the product ${\cal V}(j,q){\times}{\cal A}(j,q)$ can be taken to be momentum 
independent over the small $q$ range. 
In the long-wavelength limit, this gives the estimate:
${\cal A}(j,q) {\approx} \, { {\hbar}{\Om}_{j0} }/
({E_{g}}{ \sqrt{\,1 - {\tau}_j \,} } - { {\hbar}{\Om}_{j0} }/{E_{g}} )$.

The last element ${\cal K}(q)$ associated with the derivates of $\Delta({\bk},{\bq})$
measures at $k=0$ the effective curvature of the conductivity and valence bands.
In contrast to ${\cal V}(j,q)$ and ${\cal A}(j,q)$,
the role of the quantity ${\cal K}(q)$ is not so important
for two reasons. First, it is not directly related to the lattice properties.
Second, based on electronic structure calculations in polar materials like
$\rm ABO_{3}$ perovskite oxides, it is reasonable to assume that peculiarities
of the relevant band structure close to the Fermi level as well as features
of the Fermi surface topology do not critically influence the quantity ${\cal K}(q)$.
Therefore, for the sake of simplicity, assuming that the bands
are parabolic around $k=0$,
we restrict here to the case ${\cal K}(q){\approx}\,(1+m_{2}/m_{1})$ where 
$m_{1}$ corresponds to the bare effective mass in the valence band.
(Note that in more general case the relevant estimate of ${\cal K}$ 
may include the summation over $q$.)
Of course, it is not guaranteed that such an approximation of ${\cal K}(q)$
would be almost the same for all the $\rm ABO_{3}$ systems of interest.
However, by above considerations, the assumption of the functional non-importance
of ${\cal K}(q)$ is quite evident in the present case.

With these remarks in mind, the relatively simple expression that can be used
to find upper-bound estimates for ${\la}_{\rm TO}(j)\,$
is obtained by gathering all three observed estimates together.
The quantity ${\la}_{\rm TO}$ is then estimated in terms of observable parameters: 
\begin{equation}\label{lambda4}
{\la}_{\rm TO}(j)   \lesssim
\frac{1}{4}
\left( 1 + \frac {m_{2}} {m_{1}} \right)
\frac { {\tau}_j } { \sqrt{\,1 - {\tau}_j \,} - { {\hbar}{\Om}_{j0} }/{E_{g}} \, } \,
\frac { {\hbar}{\Om}_{j0} } {E_{g}} \, .
\end{equation}
In Eq.~(\ref{lambda4}), the adiabatic ratio ${ {\hbar}{\Om}_{j0} }/{E_{g}}$
serves as a restricting factor which minimizes significantly the magnitude
of ${\la}_{\rm TO}(j)$.
This implies that there is no need to find more sharper estimates
since the ratio of two parameters, ${\hbar}{\Om}_{j0}$ and $E_{g}$,
is sufficiently small for systems under consideration.
Thus, in order to ensure smallness of ${\la}_{\rm TO}(j)$ for any polar compound
of $\rm ABO_{3}$ perovskite oxides, 
we need only the corresponding values of material parameters such as
${\epsilon}_{\infty}$, ${S}(j)$, ${\Om}_{j0}$, and $E_{g}$
as well as the prefactor $(1+m_{2}/m_{1})$ (which can be deduced
from electronic-band-structure calculations and/or from experimental data).
Upper-bound estimates for the parameter ${\la}_{\rm TO}(j)$ calculated 
with the help of Eq.~(\ref{lambda4}) for representative
polar compounds belonging to typical perovskites are presented in Table~1.
Note that these estimates can be referred to as first principles because
Eq.~(\ref{lambda4}) contains no adjustable parameters.
\begin{table}[pt]\label{table1}
\caption{
Upper-bound estimates of ${\la}_{\rm TO}(j)$ calculated for some typical representatives
of polar perovskites and $\rm{A^{IV}B^{VI}}$ semiconductors.
The estimates employ $(1+m_{2}/m_{1})\,{\leq}\,3$
for the perovskites~\cite{Ohta1} and 
$(1+m_{2}/m_{1})\,{\leq}\,2.2\,$ for the $\rm{A^{IV}B^{VI}}$ compounds~\cite{Kaidanov}.
The rest of the material parameters were taken as in the previous work~\cite{Pishtshev1}.
}
\centering
\footnotesize
\begin{tabular}{@{}l*{10}cccccccc@{}} \toprule
Compound & ${\epsilon}_{\infty}$ & $E_{g}$ (eV) & $j$ & ${\Om}_{j0}$ & $\widetilde{S}(j)$ & ${\tau}_j$ & & ${\la}_{\rm TO}(j)$ & ${\la}_{\rm TO}=$\\
& & & &           &     &     &     &  & ${\Sigma}_{j}\,{\la}_{\rm TO}(j)$ \\
%
\colrule
${\rm SrTiO_3}$ & ${\rm 5.2}$ & ${\rm 3.3}$ & ${\rm 1}$ & ${\rm  88}$ & ${\rm 310}$ & ${\rm 0.977}$ & & $\leq {\rm 0.016}$ & $\leq {\rm 0.02}$\\
${\rm }$        & ${\rm }$    & ${\rm }$    & ${\rm 2}$ & ${\rm 176}$ & ${\rm 3.6}$ & ${\rm 0.333}$ & & $\leq {\rm 0.002}$ \\
${\rm }$        & ${\rm }$    & ${\rm }$    & ${\rm 3}$ & ${\rm 544}$ & ${\rm 1.6}$ & ${\rm 0.178}$ & & $\leq {\rm 0.003}$ \\
%
${\rm BaTiO_3}$ & ${\rm 5.2}$ & ${\rm 3.3}$ & ${\rm 1}$ & ${\rm  42}$ & ${\rm 1250}$ & ${\rm 0.994}$ & & $\leq {\rm 0.016}$ & $\leq {\rm 0.02}$\\
${\rm }$        & ${\rm }$    & ${\rm }$    & ${\rm 2}$ & ${\rm 182}$ & ${\rm 2.2}$  & ${\rm 0.234}$ & & $\leq {\rm 0.001}$ \\
${\rm }$        & ${\rm }$    & ${\rm }$    & ${\rm 3}$ & ${\rm 500}$ & ${\rm 0.8}$  & ${\rm 0.100}$ & & $\leq {\rm 0.002}$ \\
%
${\rm  KNbO_3}$ & ${\rm 4.7}$ & ${\rm 3.1}$ & ${\rm 1}$ & ${\rm  96}$ & ${\rm 230}$ & ${\rm 0.972}$ & & $\leq {\rm 0.017}$ & $\leq {\rm 0.03}$\\
${\rm }$        & ${\rm }$    & ${\rm }$    & ${\rm 2}$ & ${\rm 198}$ & ${\rm 5.5}$ & ${\rm 0.451}$ & & $\leq {\rm 0.004}$ \\
${\rm }$        & ${\rm }$    & ${\rm }$    & ${\rm 3}$ & ${\rm 521}$ & ${\rm 3.1}$ & ${\rm 0.316}$ & & $\leq {\rm 0.006}$ \\
\colrule
${\rm PbTe}$ & ${\rm 32.8}$ & ${\rm 0.31}$ & ${\rm }$ & ${\rm 32}$ & ${\rm 632}$ & ${\rm 0.948}$ & & $\leq {\rm 0.031}$ \\
${\rm SnTe}$ & ${\rm 37.0}$ & ${\rm 0.35}$ & ${\rm }$ & ${\rm 45}$ & ${\rm 330}$ & ${\rm 0.894}$ & & $\leq {\rm 0.025}$ \\
${\rm PbSe}$ & ${\rm 22.9}$ & ${\rm 0.28}$ & ${\rm }$ & ${\rm 39}$ & ${\rm 289}$ & ${\rm 0.921}$ & & $\leq {\rm 0.033}$ \\
${\rm PbS}$ & ${\rm 17.2}$ & ${\rm 0.42}$ & ${\rm }$ & ${\rm 67}$ & ${\rm 155}$ & ${\rm 0.890}$ & & $\leq {\rm 0.031}$ \\
%
\botrule
\end{tabular}
\end{table}

Based on the comparative analysis of the obtained results,
we can draw several general conclusions and characterize the corresponding
roles of the {\elTOph} interaction as follows:\\
(i) First of all, regarding effects associated with the {\elTOph} interaction,
we stress a key distinction:
On the one hand, it has been established that the large magnitude
of the {\elTOph} interaction (${\sim}\,{\rm 5}$~eV/{\AA} in the wide-gap compounds
like $\rm ABO_{3}$ perovskite oxides and ${\sim}\,{\rm 0.6}$~eV/{\AA} in
the $\rm{A^{IV}B^{VI}}$ narrow-gap semiconductors) is an important inherent property
of polar crystals~\cite{Pishtshev1,Pishtshev2}.
According to the vibronic theory (e.g.,~\cite{Kristoffel1}),
this provides a significant softening of the zone-center TO vibrations,
and may lead to a structural instability of a crystal lattice.
On the other hand, the contribution of
the el-TO-ph interaction to ${\la}$ was found to be suppressed
by the ratio ${ {\hbar}{\Om}_{j0} }/{E_{g}}$.
For example, the data in Table~1 indicate that in the $\rm ABO_{3}$ materials
the highest values of ${\la}_{\rm TO}(j)$ are in the range
of $0.016$--$0.017$ for the low-lying TO modes ($j=1$).
The high-lying modes have the smaller ${\la}_{\rm TO}(j)$ ($j=2,3$) because
the relevant el-TO-ph interactions are smaller~\cite{Pishtshev1}.
In addition to these numerical estimates, we have presented in Table 1
the corresponding results for $\rm{A^{IV}B^{VI}}$ narrow-gap semiconductors.
It is easy to see that similar estimates are also obtained for these materials.\\
(ii) The suppression of the contribution of the {\elTOph} interaction
to the total electron-phonon coupling $\la$ is needed to be clarified.
From a physical point of view, one may distinguish two factors why this is so.
Firstly, note that one of the main features of the {\elTOph} interaction
is its direct connection with the relevant changes
in the electron density distribution~\cite{Pishtshev2}.
The resulting shift of the electron density depends on interband
transitions (hybridization) which exhibit
an energy scale compared with the magnitude of $E_{g}$.
In parallel, the carrier scattering effects associated
with the long-wavelength TO phonon modes have the typical energy scale
of the order of ${\hbar}{\Om}_{j0}\,$.
Because the corresponding processes proceed in different ways,
the smallness of the contribution of the {\elTOph} interaction to ${\la}$
can be explained by the great disproportion between these scales: 
${\hbar}{\Om}_{j0}{\ll}E_{g}$.
Secondly, note that the {\elTOph} interaction directly depends on 
the electronic screening~\cite{Pishtshev2}: the factor $({\epsilon}_{\infty}+2)$
in Eq.~(\ref{lambda3A}) accounts for local-fields effects-induced partial screening
of the bare {\elTOph} interaction.
This implies that when the values of $\widetilde{S}(j)$ and
${\epsilon}_{\infty}$ are comparable, the local-fields effects are not to be ignored.\\
(iii) The strength of the Fr\"{o}hlich interaction of \nobreak electrons with the polar
long-wavelength longitudinal optical (LO) vibrations can be measured by the coupling
constant ${\alpha}_{eff}$ represented as a sum of partial constants~\cite{Devreese}:
${\alpha}_{eff}=\nobreak{\Sigma}_{i}\,{\alpha}(i)$
where the index $i$ belongs to the relevant vibrational branches.
A similar summation can also be introduced for the TO phonons as
${\la}_{\rm TO}={\Sigma}_{j}\,{\la}_{\rm TO}(j)\,$.
Since the effective contribution of the LO phonons to ${\la}$
is $\sim{\alpha}_{eff}/6$ (for weak coupling), 
the ratio ${\la}_{\rm TO}/({\alpha}_{eff}/6)$ gives then a qualitative comparison of
the corresponding contributions associated with the TO and LO phonons.
Using the polaron coupling constant ${\alpha}_{eff}{\sim}\nobreak2$ reported in~\cite{Devreese}
and the data of Table 1, one can readily see that for $\rm SrTiO_3$
the ratio ${\la}_{\rm TO}/({\alpha}_{eff}/6)\leq0.06\,$. 
As it is apparent from this comparison, the polar LO phonons
tend to provide much stronger couplings of electrons 
than the long-wavelength TO phonons.
This implies that the properties of the conduction electrons are weakly affected
by the the long-wavelength {\elTOph} interaction,
no matter how great the strength of this interaction is.\\ 
(iv) In the context of superconductivity in $\rm SrTiO_3$, it is of interest
to note that, although the {\elTOph} interaction produces little influence on ${\la}\,$,
the relative contribution of the {\elTOph} interaction might be capable of
playing some role in the doping phase diagram. (Here, we have followed the arguments
presented within a two-band scenario of superconductivity~\cite{Kristoffel3,Bussmann-Holder}.)\\
(v) Note that the derivation of Eq.~(\ref{lambda4}) assumed an adequate choice
of simplifications to recognize the relevant estimates for ${\la}_{\rm TO}$.
This allows us to test the applicability range of Eq.~(\ref{lambda4}) to other systems.
Let us demonstrate that Eq.~(\ref{lambda4})
does also work in a Mott-Hubbard system iron monosilicide ${\rm FeSi}$.
Our goal here is with the help of the experimental data relating to ${\rm FeSi}$
to estimate values of ${\la}_{\rm TO}(j)$ for the four TO zone-center phonon modes
and compare them with the corresponding values reported in~\cite{Marel2}.
Using the results of~\cite{Paschen}, we choose the prefactor $(1+m_{2}/m_{1}){\sim}\,2$
within the qualitative picture of the electronic spectra.
\begin{table}[pt]\label{table2}
\caption{
Comparison of the values of ${\la}_{\rm TO}(j)$ for ${\rm FeSi}$. The estimates were obtained
from Eq.~(\ref{lambda4}) with the use of $E_{g}={\rm 0.14}$ ${\rm eV}$ (a direct gap
value)~\cite{Paschen}, ${\epsilon}_{\infty}=8$~\cite{Bocelli}, and experimental values of
lattice TO vibration frequencies and strengths given in~\cite{Marel2}.
}
\centering
\footnotesize
\begin{tabular}{@{}l*{10}cccccc@{}} \toprule
TO mode ($j$)     & ${\rm 1}$ & ${\rm 2}$ & ${\rm 3}$ & ${\rm 4}$ &  &  \\
\colrule
${\Om}_{j0}$ (${\rm cm^{-1}}$)     & ${\rm 198}$ & ${\rm 318}$ & ${\rm 338}$ & ${\rm 445}$ & & \\
$\widetilde{S}(j)$& ${\rm 8.0}$& ${\rm 8.5}$ & ${\rm 2.8}$ & ${\rm 1.3}$ &  &  \\
${\tau}_j$& ${\rm 0.44}$& ${\rm 0.46}$ & ${\rm 0.22}$ & ${\rm 0.11}$ &  & ${\la}_{\rm TO}=$ \\
${\la}_{\rm TO}(j)$ &           &     &     &     &  & ${\Sigma}_{j}\,{\la}_{\rm TO}(j)$ \\
%
\colrule
This study &${\rm 0.068}$ &${\rm 0.142}$ &${\rm 0.056}$ &${\rm 0.041}$ && ${\rm 0.307}$\\
Ref.~\cite{Marel2}  &${\rm 0.063}$ &${\rm 0.080}$ &${\rm 0.096}$ &${\rm 0.036}$ && ${\rm 0.275}$\\
%
\botrule
\end{tabular}
\end{table}
The calculated values of ${\la}_{\rm TO}(j)$ are shown in Table 2 along with
the values obtained in~\cite{Marel2}. As expected, the agreement is fairly good.

Note that the knowledge of material parameters, such as ${\epsilon}_{\infty}$,
${S}(j)$, ${\Om}_{j0}$, and $E_{g}$, allows us to analyze the relevant interband
hybridization effects in ${\rm FeSi}$.
In Ref.~\cite{Pishtshev1} we have shown that this can be made by estimating 
the magnitude of the effective constants ${\bar{g}}^{2}(j)$
(representing the squared interband  {\elTOph} interaction averaged 
over the given electronic states). The calculated values of $\bar{g}(j)$ for ${\rm FeSi}$
are given in Table 3. 
\begin{table}[pt]\label{table3}
\caption{
Values of the {\elTOph} interaction constant $\bar{g}(j)$ and the transverse effective
charge $|Z^{*}(j)|$ provided for each zone-center polar TO vibrational mode of
${\rm FeSi}$. The estimates were made as described previously~\cite{Pishtshev1}
with the use of the material parameters of Table~2.
Two columns were added to allow comparison of results between
different compounds. The entries in the columns labeled
{\textquoteleft}(A){\textquoteright} and {\textquoteleft}(B){\textquoteright} refer to 
the estimates for ${\rm FeO}$ and ${\rm PbTe}$~\cite{Pishtshev1}, respectively.
}
\centering
\footnotesize
\begin{tabular}{@{}l*{10}cccccc@{}} \toprule
TO mode ($j$)                   & ${\rm 1}$    & ${\rm 2}$    & ${\rm 3}$    & ${\rm 4}   $ && (A) & (B)\\
\colrule
$\bar{g}(j)$ (~eV/{\AA})& ${\rm 0.27}$ & ${\rm 0.45}$ & ${\rm 0.28}$ & ${\rm 0.25}$ && ${\rm 1.9}$ & ${\rm 0.65}$ \\
$|Z^{*}(j)|$                    & ${\rm 1.63}$ & ${\rm 2.70}$ & ${\rm 1.65}$ & ${\rm 1.47}$ && ${\rm 2.19}$ & ${\rm 8.37}$ \\
%
\botrule
\end{tabular}
\end{table}
It is easy to see that the values of $\bar{g}(j)$ for both narrow-gap compounds appear
to be close to each other.
In materials where the {\elTOph} interaction is operative, it can be verified
by inspection of the magnitude of the zone-center polar TO vibrational mode effective
charges $|Z^{*}(j)|$~\cite{Pishtshev2,Pishtshev1}.
This point is well illustrated in Table 3, which presents the estimated values of the transverse
effective charges, $|Z^{*}(j)|$, both for wide-gap (${\rm FeO}$) and narrow-gap (${\rm FeSi}$,
${\rm PbTe}$) compounds.
Thus, the estimates presented in Table 3 suggest that the interband channels of electronic
scatterings associated with the zone-center TO phonons play a certain role
in the lattice-dynamical properties of ${\rm FeSi}$.
%
\section{Conclusion}
In the present paper, motivated by the need to extend the results of the previous
works~\cite{Pishtshev1,Pishtshev2}, we have analyzed the influence of the interaction
of electrons with infrared-active long-wavelength TO lattice vibrations on the total
electron-phonon coupling constant $\la\,$.
To examine in detail in what extent {\elTOph} interactions are important for
the total strength of electron-phonon coupling we considered the generic
two-band model involving the interband hybridization by the TO vibrational modes. 
The main result of the analysis points out the marginal role played by
the {\elTOph} interaction on electron pairing in doped polar insulating systems:
the presented analytical and numerical arguments show that 
the partial contribution from the {\elTOph} interaction to $\la\,$ 
is much smaller than the relevant contribution from the interaction with 
the polar long-wavelength LO phonons.
The corresponding explanation is that the important physical properties of
the {\elTOph} interaction are described by the matrix elements
which mix the electron states of the valence and conduction bands separated
by relatively large energy gap.
For given intra- and interband interactions coexisting in a doped insulator,
due to distinct energy scales, the amount of the contribution of the {\elTOph}
interaction is ruled by the adiabatic ratio ${\hbar}{\Om}_{0}/{E_{g}}\,$.
In most systems of interest, the quantities, ${E_{g}}$ and ${\hbar}{\Om}_{j0}$,
are of a different order of magnitude so that the real impact of the {\elTOph} interaction
on ${\la}$ turns out to be strongly suppressed.
This conclusion has a clear physical meaning: upon completion of the intraband scatterings,
a conduction electron becomes already nearly ``undressed''.
Our studies suggest that the {\elTOph} interaction (via valence electrons) is of great concern
for the lattice and polarizability properties, the changes in the electron density distribution,
and the renormalization (soft behavior) of the zone-center TO vibrational modes.
At the same time, in the context of superconductivity in doped polar insulating compounds,
the el-TO-ph interaction due to a difference in energy scales of underlying interactions tends
to contribute little to the total effective coupling constant.
\vspace{1.cm}
\section*{Acknowledgments}
The work was supported by the ETF grant No. 6918.
%
%

\end{document}